\documentclass[conference]{IEEEtran}
\IEEEoverridecommandlockouts
\usepackage{amsmath,amsfonts}
\usepackage{algorithmic}
\usepackage{algorithm}
\usepackage{array}
\usepackage[caption=false,font=normalsize,labelfont=sf,textfont=sf]{subfig}
\usepackage{textcomp}
\usepackage{stfloats}
\usepackage{url}
\usepackage{verbatim}
\usepackage{graphicx}
\usepackage{cite}
\usepackage{siunitx}
\usepackage{comment}
\usepackage{balance}
\usepackage{color}
\newcommand{\revision}[1]{{\color{black}{#1}}}
\usepackage[normalem]{ulem}

\usepackage{multirow}
\usepackage[table,xcdraw]{xcolor}

\begin{document}


\title{Exploring Human Response Times to\\ Combinations of Audio, Haptic, and Visual Stimuli\\ from a Mobile Device\\

\thanks{This work was supported in part by a grant from the Precision Health and Integrated Diagnostics Center at Stanford and the National Science Foundation Graduate Fellowship Program.}
\thanks{\textsuperscript{$\star$}These authors contributed equally to this work.}
} 



\author{\IEEEauthorblockN{Kyle T.\ Yoshida\textsuperscript{$\star$}}
\IEEEauthorblockA{\textit{Mechanical Engineering,} \\
\textit{Stanford University}\\
Stanford, CA, USA \\
kyle3@stanford.edu\vspace{-40 pt}
}
\and
\IEEEauthorblockN{Joel X.\ Kiernan\textsuperscript{$\star$}}
\IEEEauthorblockA{\textit{Mechanical Engineering,} \\
\textit{Stanford University}\\
Stanford, CA, USA \\
joelx@stanford.edu\vspace{-40 pt}}
\and
\IEEEauthorblockN{Allison M.\ Okamura}
\IEEEauthorblockA{\textit{Mechanical Engineering,} \\
\textit{Stanford University}\\
Stanford, CA, USA \\
aokamura@stanford.edu\vspace{-40 pt}}
\and
\IEEEauthorblockN{Cara M.\ Nunez}
\IEEEauthorblockA{\textit{Sibley School of Mechanical and} \\
\textit{Aerospace Engineering, Cornell University}\\
Ithaca, NY, USA \\
cmn97@cornell.edu\vspace{-20 pt}}

}




\maketitle

\begin{abstract}

Auditory, haptic, and visual stimuli provide alerts, notifications, and information for a wide variety of applications ranging from virtual reality to wearable and hand-held devices. Response times to these stimuli have been used to assess motor control and design human-computer interaction systems. In this study, we investigate human response times to 26 combinations of auditory, haptic, and visual stimuli at three levels (high, low, and off). We developed an iOS app that presents these stimuli in random intervals and records response times on an iPhone 11. We conducted a user study with 20 participants and found that response time decreased with more types and higher levels of stimuli. The low visual condition had the slowest mean response time \revision{(mean $\pm$ standard deviation, 528 $\pm$ 105~ms)} and the condition with high levels of audio, haptic, and visual stimuli had the fastest mean response time \revision{(320 $\pm$ 43~ms)}. This work quantifies response times to multi-modal stimuli, identifies interactions between different stimuli types and levels, and introduces an app-based method that can be widely distributed to measure response time. Understanding \revision{preferences and response times for stimuli can provide insight into designing devices for} human-machine interaction.


\end{abstract}

\begin{IEEEkeywords}
Response Time, Haptics, Vibration, Auditory, Audio, Visual, Mobile, Smartphone
\end{IEEEkeywords}

\vspace{-5pt}

\section{Introduction}
Auditory, haptic, and visual stimuli are commonly used to provide alerts and notifications in daily life. Response times to these stimuli have been used to characterize human sensorimotor systems~\cite{delmas2018motor}, \revision{quantify sleep quality}~\cite{ferguson2011performance}, and design collision avoidance systems~\cite{belz1999new}. When designing human interfaces, understanding response times to auditory, haptic, and visual stimuli is vital. In some situations, engineers focus on minimizing response time to provide fast alerts for vehicle operators~\cite{belz1999new, petermeijer2017, scott2008comparison}. In other situations, engineers focus on combining stimuli to elicit responses to constraints in teleoperation or surgical robotic systems~\cite{schleer2020augmentation}. 

\begin{figure}
	\centering
	\includegraphics[width=0.85\columnwidth]{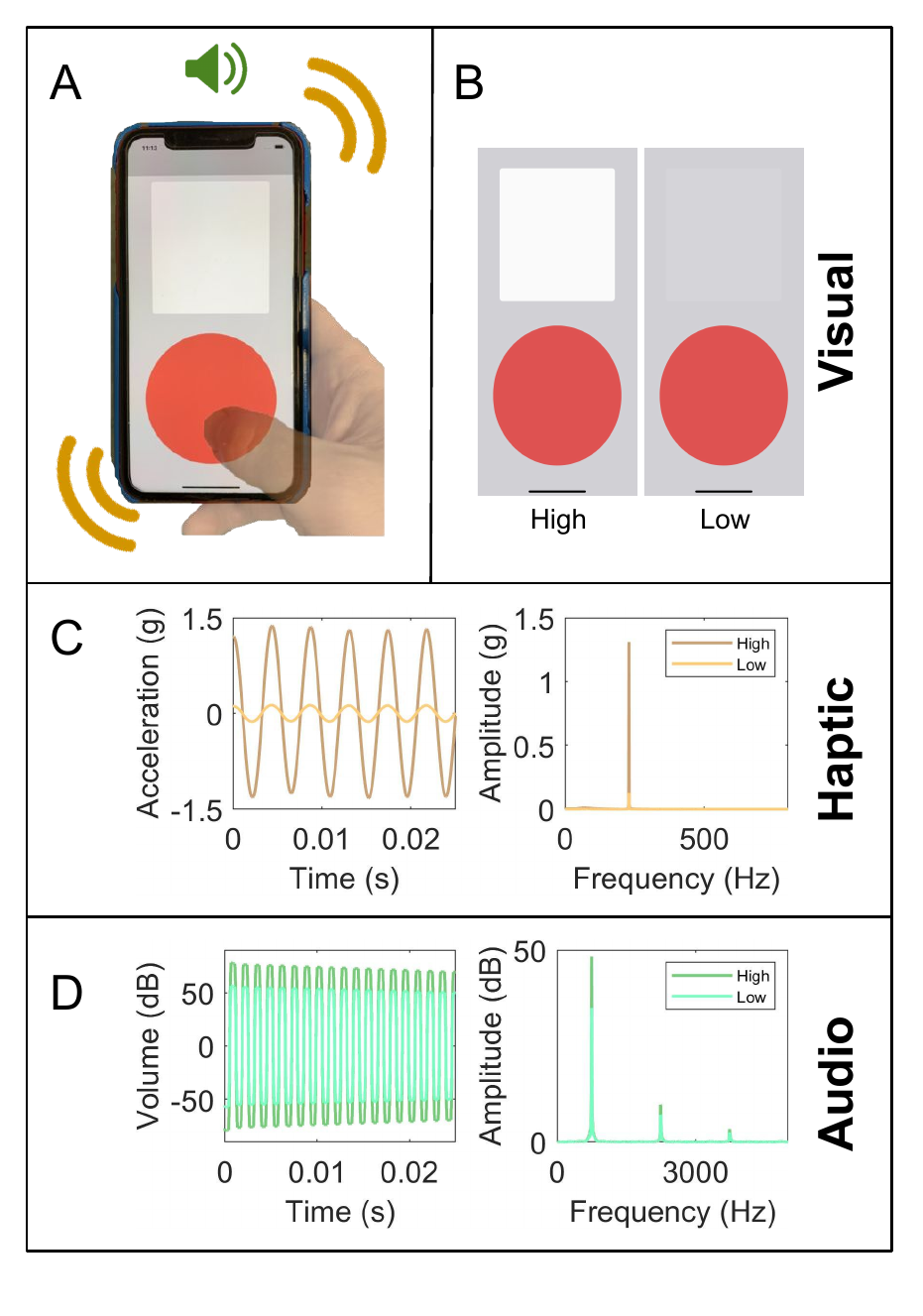}
	\vspace{-0.25in}
	\caption{Smartphone app and stimuli characteristics. We used a custom iOS app (A) to measure response times to 26 combinations of audio, haptic, and visual stimuli at three different intensity levels -- high, low, and off. \revision{The visual stimulus was a shape of varying transparency (B), the haptic stimulus was a vibration of varying intensity (C), and the audio stimulus was a tone of varying volume (D).} }\label{fig:setup}
   \vspace{-0.2 in}
\end{figure}

\revision{Prior} work on quantifying response times \revision{was} specific to a particular system and \revision{examined only} one or two types of stimuli. Peon and Prattichizzo compared response times to audio, visual, and haptic stimuli with a kinesthetic haptic device and found that response time is faster for haptic stimuli, followed by auditory and visual stimuli~\cite{peon_reactiontime}. Scott and Gray also found that subjects responded fastest to haptic stimuli, followed by auditory and visual stimuli, in a driving collision avoidance context~\cite{scott2008comparison}. In a similar trucking collision avoidance system, Belz et al. found that a combination of visual and auditory stimuli decreased mean response time, compared to standalone visual or auditory stimuli~\cite{belz1999new}. Jia et al. found similar results: a combination of visual and auditory stimuli decreases response time and increases accuracy~\cite{jia2019}. When a subject is distracted, Petermeijer et al. found that auditory and haptic stimuli elicit the fastest response times~\cite{petermeijer2017}. Instead of focusing on response times, Hecht et al.~\cite{hecht_reiner_2008} focused on human perception of combinations of stimuli. They found that when multiple stimuli are played together, subjects mainly perceive visual stimuli, and are less likely to realize that they also received a haptic or auditory stimulus. While most studies applied haptic stimuli to subjects' hands, Chan and Ng discovered that haptic stimulus location while seated does not affect response times~\cite{chan_2012}.

\revision{Prior work mainly focuses on response speeds of standalone stimuli. Peon and Prattichizzo~\cite{peon_reactiontime}, Chan and Ng~\cite{chan_2012}, and Scott and Gray~\cite{scott2008comparison} quantified the response times of standalone auditory, visual, or haptic stimuli, but not their combinations. Peon and Prattichizzo~\cite{peon_reactiontime} also examined how high and low levels of auditory and vibrotactile stimuli impacted response time. Jia and Shi~\cite{jia2019} and Belz et al.~\cite{belz1999new} investigated how the presentation of both visual and auditory stimuli impact response time. Hecht et al.~\cite{hecht_reiner_2008} did not measure response times, but they did find that visual stimuli are the most likely to be identified when played in combination with other stimuli.

}


In this work, we aim to understand and quantify the effect that the interactions between auditory, haptic and visual stimuli have on response time. We present our methods using a phone-based system in Section II. In Section III, we discuss our results. In Section IV, we discuss implications of the results, followed by our conclusion and directions for future work in Section V.


\section{Methods}
\subsection{Smartphone App}
\begin{figure}
	\centering
	\includegraphics[width=0.9\columnwidth]{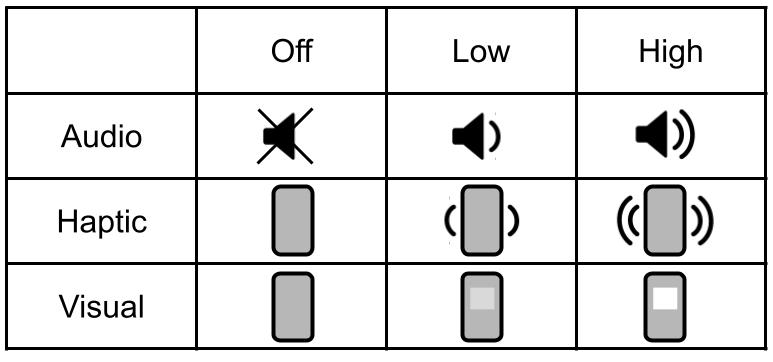}
	\vspace{-0.15in}
	\caption{Table of stimuli combinations. Each stimulus is paired with any of the other possible stimuli in the other rows, with the exception of the no audio/no haptic/no visual stimulus condition. }
 \label{fig:method}
 		    \vspace{-0.15in}

\end{figure}

We designed an iOS app (Swift, v5.7.1; XCode v14.1; https://www.github.com/charm-lab/reaction\_time\_app) for an Apple iPhone 11 that presents auditory, visual, and haptic stimuli in randomly-spaced 3-6~s intervals while recording user response times via a button press (Fig.~\ref{fig:setup}). Each stimulus was tested at three levels -- off, low, and high. All combinations and levels of these stimuli were tested, except for the case where all stimuli are off (as this would elicit no measurable response) (Fig.~\ref{fig:method}). The app plays each of these 26 combinations a total of 5 times over a 13-minute duration. Each combination is played once before any repetitions.

The auditory stimulus is a tone (peak frequency at 746 hz) played at either high (90 dB) or low (65 dB) (Fig.~\ref{fig:setup}). The tone was taken from a pre-recorded soundtrack~\cite{audio_soundtrack}, and played using the AVF Audio framework (Apple Inc.) with an \revision{\textit{AVAudioPlayer}
} volume of 0.1 and 0.01. The output volume was measured using a sound meter (Splend Apps) on the Samsung Galaxy S8 at the output of the speaker of an iPhone 11 when the phone volume is on its maximum setting. The haptic stimulus is created using the Core Haptics Framework (Apple Inc.), which can be used to play a discrete pulse by modifying Core Haptics variables of \revision{\textit{hapticIntensity} or \textit{hapticSharpness}
~\cite{CoreHaptics}}. The haptic stimulus is either a high amplitude vibration (\revision{\textit{hapticIntensity} = 1.0, \textit{hapticSharpness} = 1.0) or low amplitude vibration (\textit{hapticIntensity} = 0.3, \textit{hapticSharpness} = 1.0) that plays for 0.1~s. }
Vibrations were measured using an accelerometer (Analog Devices, EVAL-ADXL354CZ) \revision{and }
a DAQ (National Instruments, NI9220), \revision{interfaced with} Matlab (Mathworks)~\cite{adenekan_phone}. Both vibrations have a predominant frequency of 230~Hz. The low vibration has an amplitude of 0.12~g, and the high vibration has an amplitude of 1.3~g (Fig.~\ref{fig:setup}). The visual stimulus is a white square (4.7~cm $\times$~4.7 cm) that appears against a grey background for 0.5~s. The square appears with either high transparency (\revision{\textit{Opacity}} 
= 0.1) or low transparency (\revision{\textit{Opacity}} 
= 0.9) (Fig.~\ref{fig:setup}). The values for the levels were chosen based off internal pilot studies such that the low levels would be as small as possible while still perceivable to all subjects and the high levels were as high as possible. 

\subsection{User Study}

We conducted a user study to measure the response times to the different combinations and levels of auditory, visual, and haptic stimuli by using our app. Participants were pre-screened, to select only those who were right-handed and over the age of 18. The experimental protocol was approved by \revision{the Stanford University }
Institutional Review Board, and all participants gave informed consent. 

\revision{Our user study included 20 participants (13 female, 6 male, 1 non-binary; aged 20-29). None of the users reported familiarity with human-machine interaction devices, neurological conditions, or injuries. Data from a total of 24 participants was collected, but four were removed. Two were removed due to excessive missed responses resulting in incomplete data that could not be analyzed, and two were outliers with all data outside the IQR method of outlier detection~\cite{outlierdetectionberger2021}. }

\revision{Participants first} completed a pre-survey to gather demographic information and \revision{reported} their experience with computer-machine interfaces. 
Next, each participant completed three 13-minute blocks of data collection which consisted of 5 trials of each of the 26 audio, visual, and haptic combinations. Therefore, in total, there were 15 response times collected for each stimuli combination and a total of 390 response times per participant.

Between each block of data collection, participants \revision{rested for 3 minutes}. \revision{During data collection periods, p}articipants held the phone in their right hand and could rest their arm on a table during the study. The phone had a custom phone case to normalize finger placement across the back of the phone~\cite{yoshida2023cognitive}. Participants were also instructed to not touch the phone with their left hand during the experiment and to hold their right thumb over the input button to minimize their response time. Participants were also instructed that the stimuli would occur at random intervals, so they should not try to predict when the stimuli would occur. After the experiment, participants completed a post-survey, in which they ranked their preference and perceived speed for responding to the audio, haptic, and visual stimuli in addition to providing \revision{additional} feedback about the experiment.

\subsection{Statistical Analysis}

Response times were measured as the time between the onset of the visual, haptic, and/or audio stimulus and the onset of the participant's button press. Response times shorter than 10~ms and greater than 1500~ms were treated as false positives and removed from further analysis. These were removed because they were either faster than humanly possible~\cite{MOLHOLM2002115} or \revision{were too slow and more likely to be an} accidental tap disconnected from the presented stimulus. If a participant missed any of the stimuli, those individual responses were omitted from the participant's average for that particular stimulus combination. If a participant missed all the occurrences for a particular stimuli, the participant was removed. 

A 3-way repeated measures ANOVA was used to identify main effects or interactions when all three stimuli were played. Three 2-way repeated measures ANOVAs were used to identify main effects or interaction when only two stimuli were played in the absence of the third. Lastly, a 2-way repeated measures ANOVA was used with the stimulus type and stimulus level as main factors for all the data during which only one stimulus is played. Greenhouse-Geisser corrections were applied to within-subjects factors if sphericity was violated. ANOVAs were followed by post-hoc T-tests with Bonferroni corrections. Friedman's Tests were used to compare results from the survey. R Studio~\cite{rstudio} was used for statistical analysis along with the tidyverse package~\cite{Wickham2019-mu}. The significance threshold was set at $p<\num{0.05}$. All plots were created in Matlab (Mathworks).


\begin{figure*}
	\centering
	\includegraphics[width=1\textwidth]{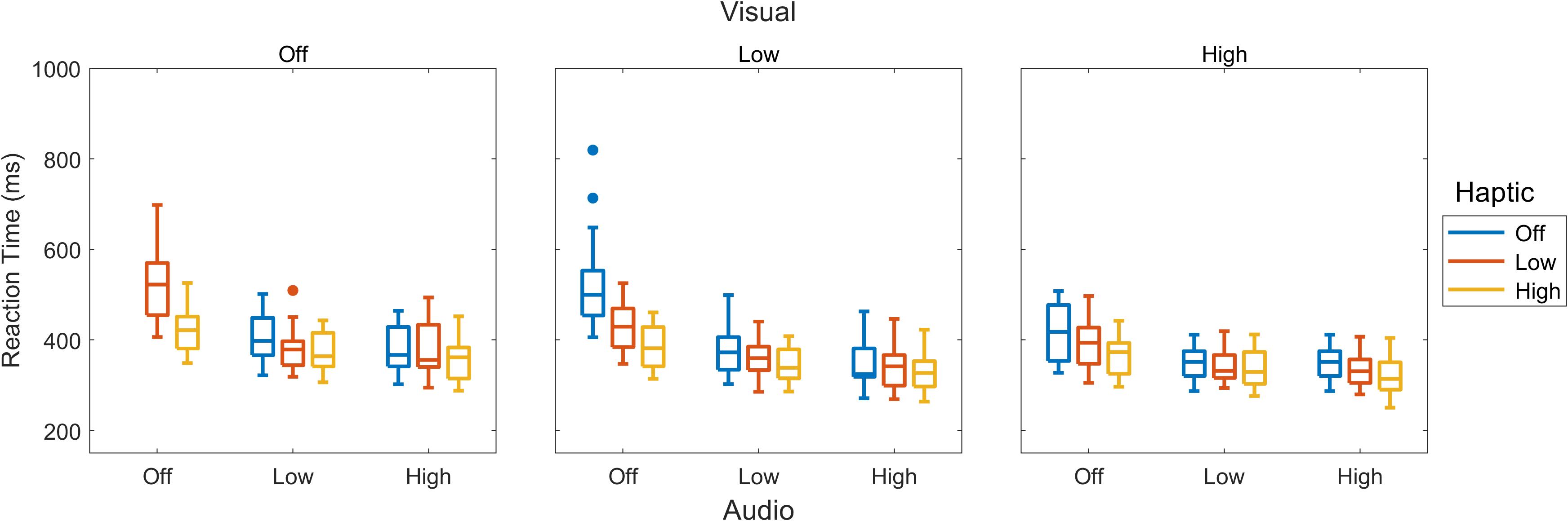}
	\vspace{-0.2in}
	\caption{Boxplots showing the response times for each combination of audio, haptic, and visual stimuli at three levels (off, low, high). Visual levels are denoted by subplots, haptic levels are denoted by color, and audio levels are denoted by x-axis location. The mean response time is the fastest in the condition with high levels of all three stimuli (\revision{mean $\pm$ standard deviation, }320 $\pm$ 43~ms) and slowest in the low visual condition (528 $\pm$ 105~ms).}\label{fig:alldata}
		   \vspace{-0.15in}
\end{figure*}

\begin{figure}
	\centering
	\includegraphics[width=\columnwidth]{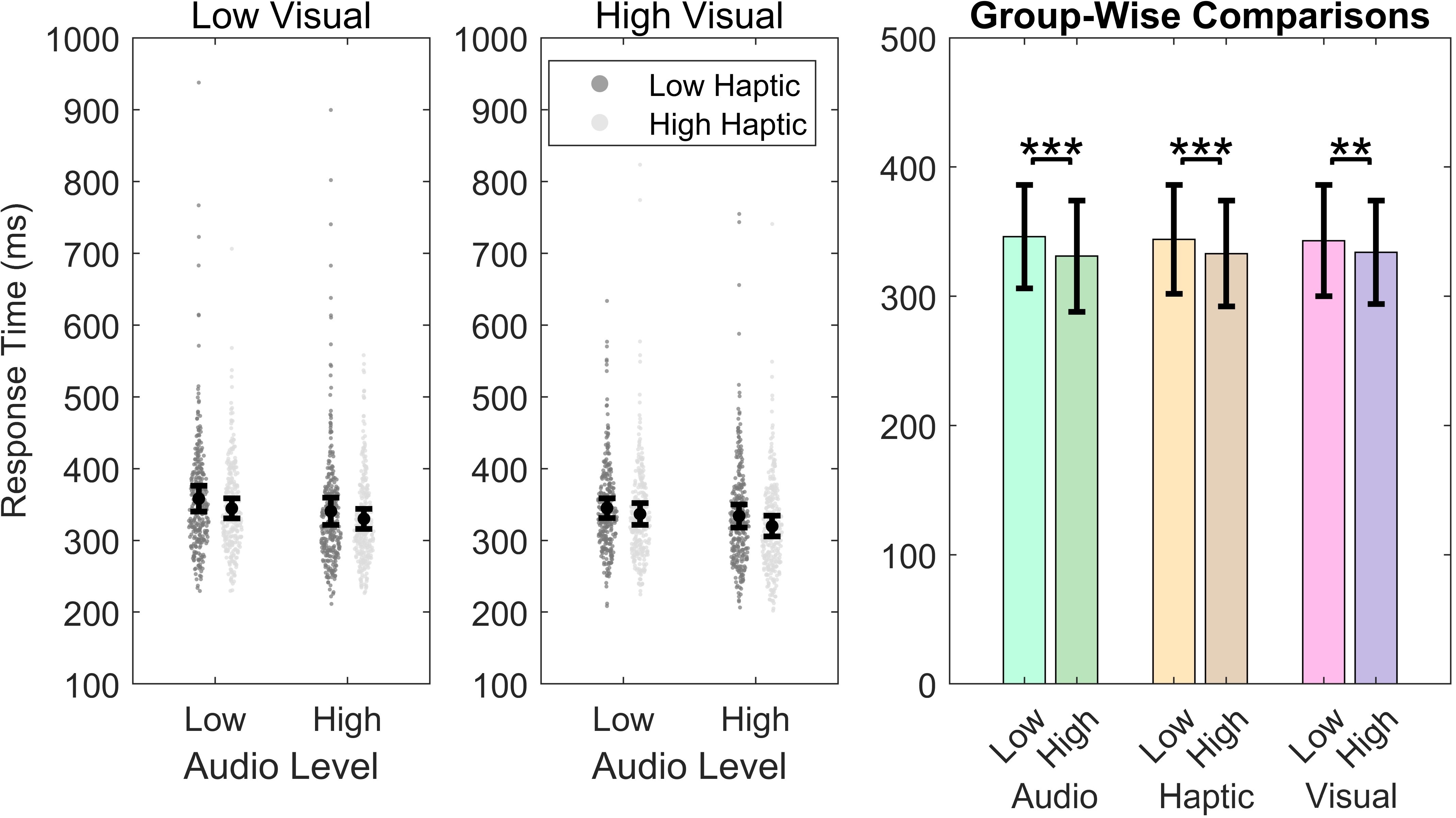}
	\vspace{-0.25in}
	\caption{\revision{Scatter plots showing tri-modal response times for each trial superimposed with the mean and standard error 
 (left)}. \revision{Means and standard errors for grouped p}ost-hoc pairwise comparisons showing the impact of level for each stimulus (right). \revision{Standard significance notation is used for p-values ($**: 0.001 < p < 0.01$, $*\!*\!*: p < 0.001$). }When all three stimuli are delivered simultaneously, response time is modulated by the level of each stimulus.  }\label{fig:3way}
     \vspace{-0.15 in}

\end{figure}


\section{Results}

Figure~\ref{fig:alldata} shows boxplots of response times across all 26 conditions. The fastest mean response time recorded for a particular subject was 250~ms, and the slowest was 819~ms. The slowest mean response time occurred for the low visual condition \revision{(mean $\pm$ standard deviation, 528 $\pm$ 105~ms)}, and the fastest mean response time occurred for the condition with high levels of all three stimuli (320 $\pm$ 43~ms). The overall mean response time across all conditions was (380 $\pm$ 52~ms). 

\subsection{\revision{Tri-Modal Stimuli
}}
Figure~\ref{fig:3way} displays the mean response time for all conditions in which all stimuli (audio, haptic, and visual) were present with two different levels (high/low). The mean response time was fastest in the condition with high levels of all three stimuli (320 $\pm$ 43~\revision{m}s) and slowest in the condition with low levels of all three stimuli (358 $\pm$ 43~\revision{m}s). A three-way repeated measures ANOVA with audio, haptic, and visual as independent factors with two levels of each (high/low) revealed a main effect of audio ($F(1,19)=28.1$, $p=\num{4.07E-5}$, $\eta_{p}^{2}=0.597$), haptic ($F(1,19)=23.8$, $p=\num{1.05E-4}$, $\eta_{p}^{2}=0.556$), and visual ($F(1,19)=9.01$, $p=\num{7.00E-3}$, $\eta_{p}^{2}=0.322$) stimuli on response time. These main effects were not qualified by any interactions between audio and haptic stimuli ($F(1,19)=0.155$, $p=\num{6.98E-1}$, $\eta_{p}^{2}=0.008$), audio and visual stimuli ($F(1,19)=0.129$, $p=\num{7.23E-1}$, $\eta_{p}^{2}=0.007$), haptic and visual stimuli ($F(1,19)=0.039$, $p=\num{8.45E-1}$, $\eta_{p}^{2}=0.002$), or all three stimuli ($F(1,19)=2.559$, $p=\num{1.26E-1}$, $\eta_{p}^{2}=0.119$). Post-hoc pairwise comparisons showed that a higher level of \revision{an} audio ($p=\num{4.07E-5}$), haptic ($p=\num{1.05E-4}$), or visual ($p=\num{7.33E-3}$) stimulus result\revision{s} in a faster response time. The significant main effects and post-hoc comparisons mean that when all three stimuli are delivered, the response time will change depending on the strength of each stimulus.

\begin{figure}
	\centering
	\includegraphics[width=\columnwidth]{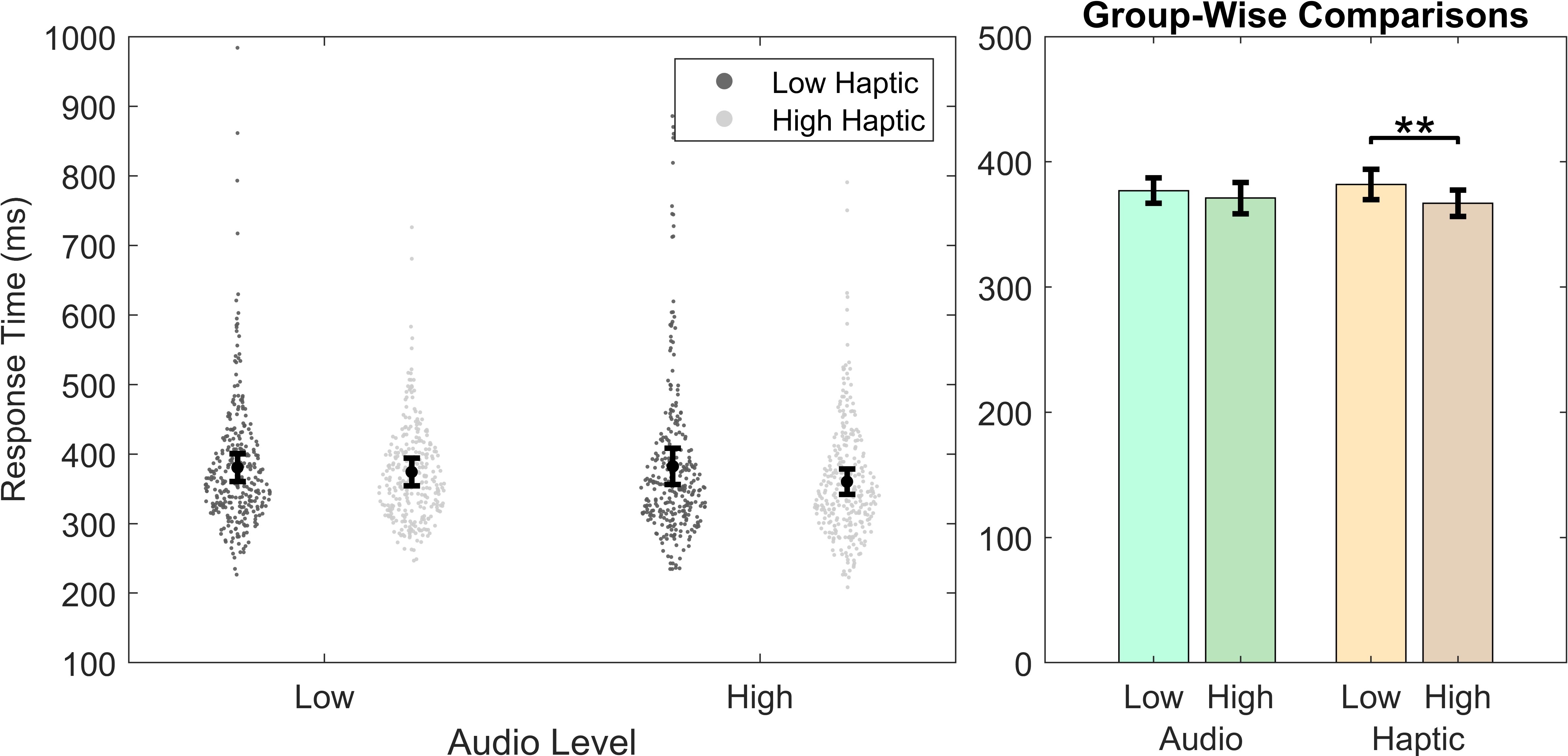}
	\vspace{-0.25in}
	\caption{Scatter plot showing response times for all trials for which audio and haptic stimuli are simultaneously delivered (left). \revision{Response times for individual trials are shown in the scatter plot, superimposed with the mean and standard error. Means and standard errors for grouped p}ost-hoc pairwise comparisons showing the impact of the level of each stimulus (right). \revision{Standard significance notation is used for p-values ($**: 0.001 < p < 0.01$). }\revision{
  When audio and haptic stimuli are simultaneously delivered, the change in response time is primarily impacted by the level of the haptic stimulus. }}\label{fig:2waya}
 \vspace{-0.1 in}

\end{figure}

\begin{figure}
	\centering
	\includegraphics[width=\columnwidth]{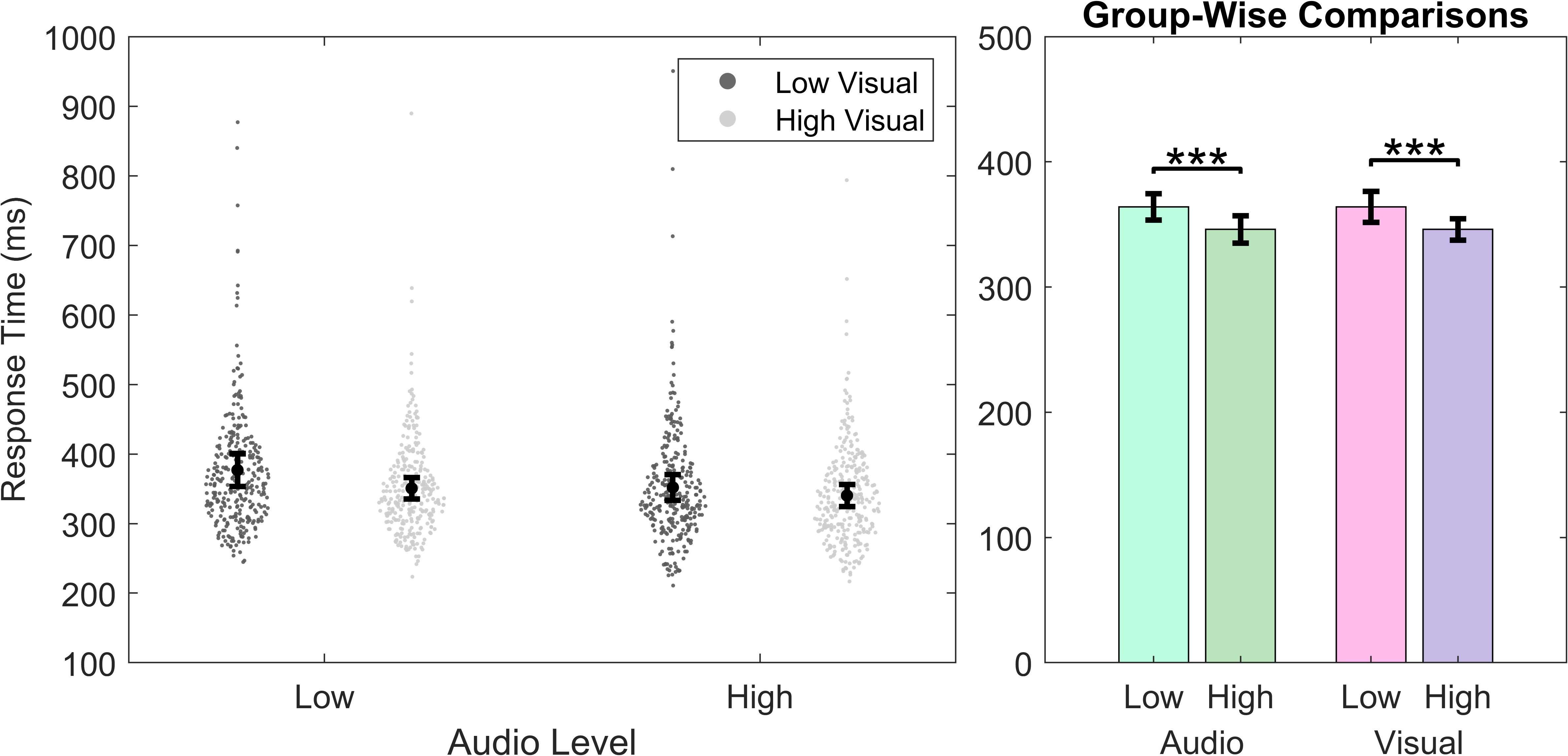}
	\vspace{-0.25in}
	\caption{Scatter plot showing response times for all trials for which audio and visual stimuli are simultaneously delivered (left). \revision{Response times for individual trials are shown in the scatter plot, superimposed with the mean and standard error. Means and standard errors for grouped p}ost-hoc pairwise comparisons showing the impact of the level of each stimulus (right). \revision{Standard significance notation is used for p-values ($*\!*\!*: p < 0.001$). }Both audio and visual stimuli have significant main effects, and when they are simultaneously delivered, the change in response time is dependent on the level of each stimulus. }\label{fig:2wayb}
    \vspace{-0.1 in}

\end{figure}

\begin{figure}
	\centering
	\includegraphics[width=\columnwidth]{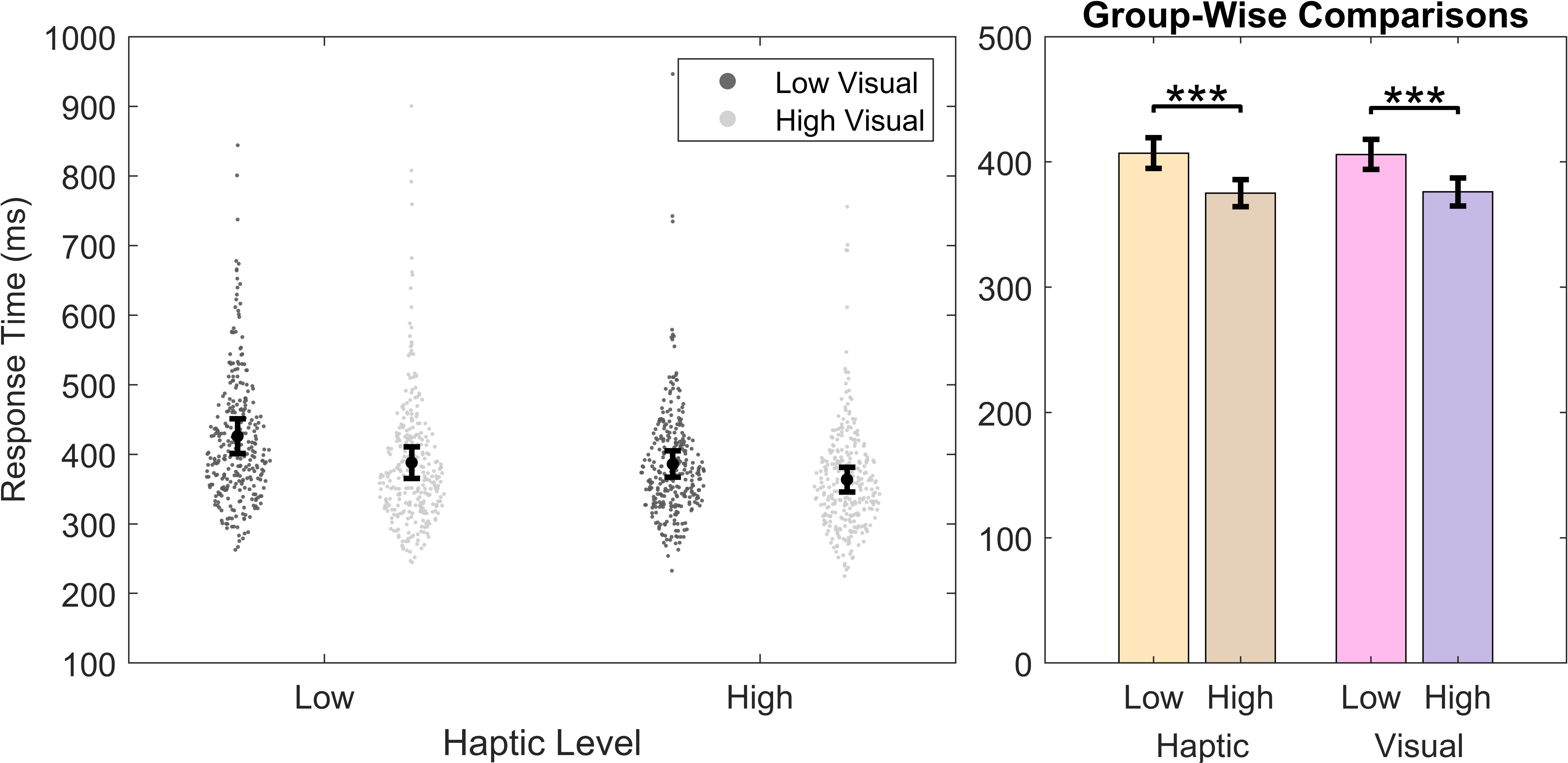}
	\vspace{-0.25in}
	\caption{Scatter plot showing response times for all trials for which haptic and visual stimuli are simultaneously delivered (left). \revision{Response times for individual trials are shown in the scatter plot, superimposed with the mean and standard error. Means and standard errors for grouped p}ost-hoc pairwise comparisons showing the impact of the level of each stimulus (right). \revision{Standard significance notation is used for p-values ($*\!*\!*: p < 0.001$).} Both haptic and visual stimuli have significant main effects, and when they are simultaneously delivered, the change in response time is dependent on the level of each stimulus. }\label{fig:2wayc}
    \vspace{-0.1 in}

\end{figure}

\subsection{\revision{Bi-Modal Stimuli
}}
We conducted three separate 2-way repeated measures ANOVAs to understand the main effects and interaction effects of audio, haptic, and visual stimuli when only two of the three are present. When only two stimuli are present, response time is the fastest with high levels of audio and haptic stimuli (341 $\pm$ 42~ms) and the slowest in the condition with low levels of haptic and visual stimuli (426 $\pm$ 53~ms). 

\subsubsection{Audio and Haptic}
Figure~\ref{fig:2waya} shows the response time of all trials for all subjects for all combinations of audio and haptic stimuli at two levels (high/low) in the absence of visual stimuli. A two-way repeated measures ANOVA with stimulus types as independent factors (audio/haptic) with two levels (high/low) revealed a main effect of haptic ($F(1,19)=11.0$, $p=\num{4.00E-5}$, $\eta_{p}^{2}=0.367$), but not audio ($F(1,19)=0.801$, $p=\num{0.382}$, $\eta_{p}^{2}=0.040$) stimuli on response time. These main effects were not qualified by any interactions between audio and haptic stimuli ($F(1,19)=3.28$, $p=\num{0.086}$, $\eta_{p}^{2}=0.147$). Post-hoc pairwise comparisons showed that a high haptic stimulus resulted in a faster response time ($p=0.003$). This means that when audio and haptic stimuli are simultaneously delivered, the change in response time depends \revision{primarily on the strength of the haptic stimulus.}

\subsubsection{Audio and Visual}
Figure~\ref{fig:2wayb} shows the response times for all combinations of audio and visual stimuli at two levels (high/low) in the absence of haptic stimuli. A two-way repeated measures ANOVA with audio and visual stimuli as independent factors with two levels (high/low) revealed a main effect of audio ($F(1,19)=20.1$, $p=\num{2.54E-4}$, $\eta_{p}^{2}=0.514$) and visual ($F(1,19)=13.4$, $p=\num{0.002}$, $\eta_{p}^{2}=0.415$) stimuli on response time. These main effects were not qualified by any interaction between audio and visual stimuli ($F(1,19)=3.97$, $p=\num{0.061}$, $\eta_{p}^{2}=0.173$). Post-hoc pairwise comparisons showed that a high audio stimulus ($p=\num{7.53E-5}$) and a high visual stimulus ($p=\num{2.09E-4}$) resulted in faster response times, meaning that when both are played, the level of each one can influence response time.

\subsubsection{Haptic and visual}
Figure~\ref{fig:2wayc} shows the response times for all combinations of haptic and visual stimuli at two levels (high/low) in the absence of audio stimuli. A two-way repeated measures ANOVA with haptic and visual stimuli as independent variables with two levels (high/low) revealed a main effect of haptic stimuli ($F(1,19)=34.8$, $p=\num{1.11E-5}$, $\eta_{p}^{2}=0.647$) and visual stimuli ($F(1,19)=49.0$, $p=\num{1.11E-6}$, $\eta_{p}^{2}=0.721$) on response time. These main effects were not qualified by an interaction between haptic and visual stimuli ($F(1,19)=2.28$, $p=\num{0.148}$, $\eta_{p}^{2}=0.107$). Post-hoc pairwise comparisons showed that a high haptic stimulus ($p=\num{4.55E-7}$) and a high visual stimulus ($p=\num{2.28E-7}$) resulted in faster response times, meaning that when both are played, the level of each one can influence response time.

\begin{figure}
	\centering
	\includegraphics[width=\columnwidth]{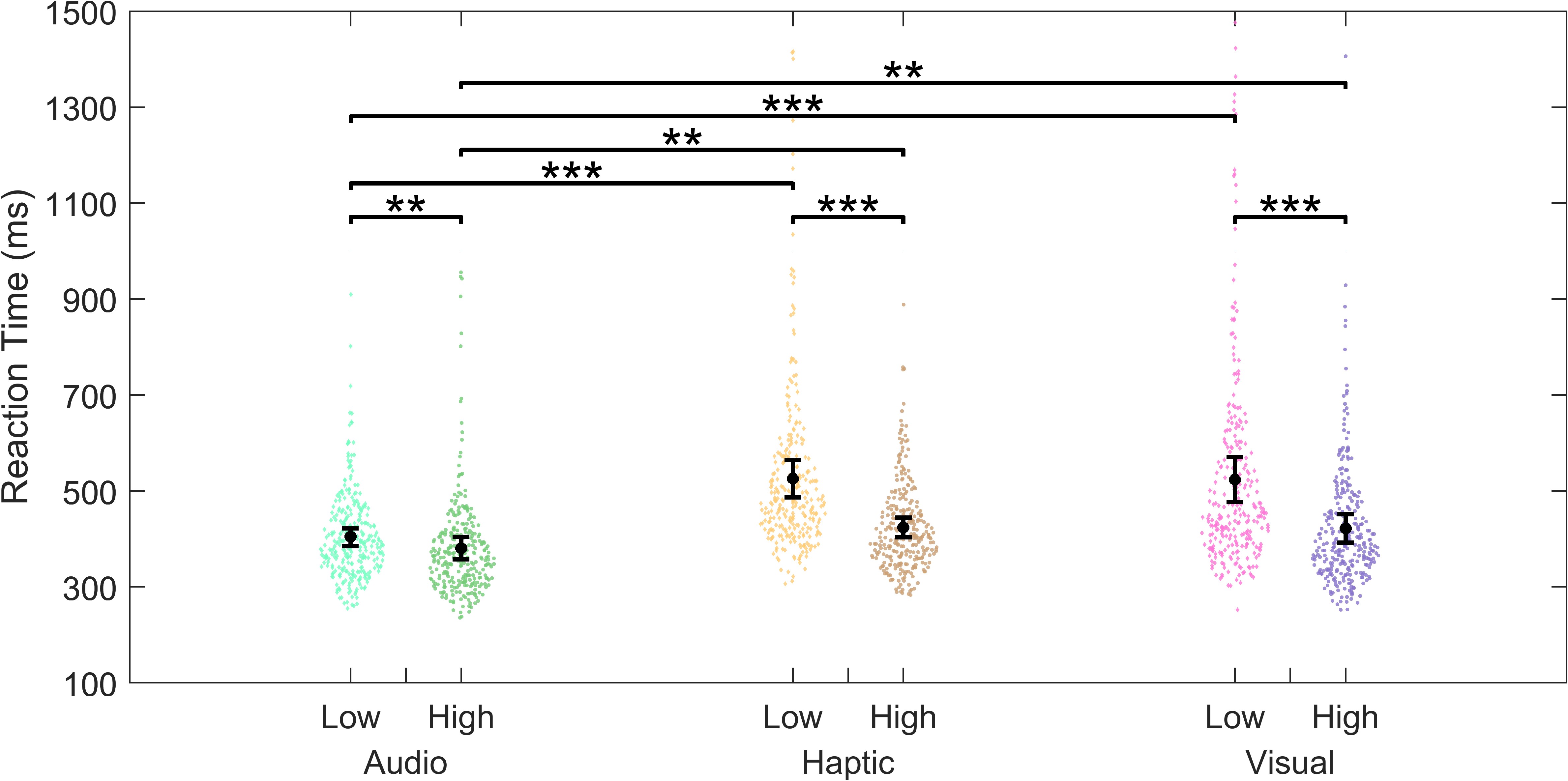}
	\vspace{-0.25in}
	\caption{\revision{Individual response times superimposed with group means and standard errors from uni-modal stimuli. Statistical significance was determined using post-hoc pairwise comparisons with p-values indicated with standard significance notation ($**: 0.001 < p < 0.01$, $*\!*\!*: p < 0.001$). }There were significant differences between the response times for the high and low levels for each stimulus. The low haptic and visual stimuli had slower response times than the low audio stimulus, and the high haptic and visual stimuli had slower response times to the high audio stimulus.}\label{fig:2waysingle}
     \vspace{-0.1 in}

\end{figure}

\begin{figure}
	\centering
	\includegraphics[width=\columnwidth]{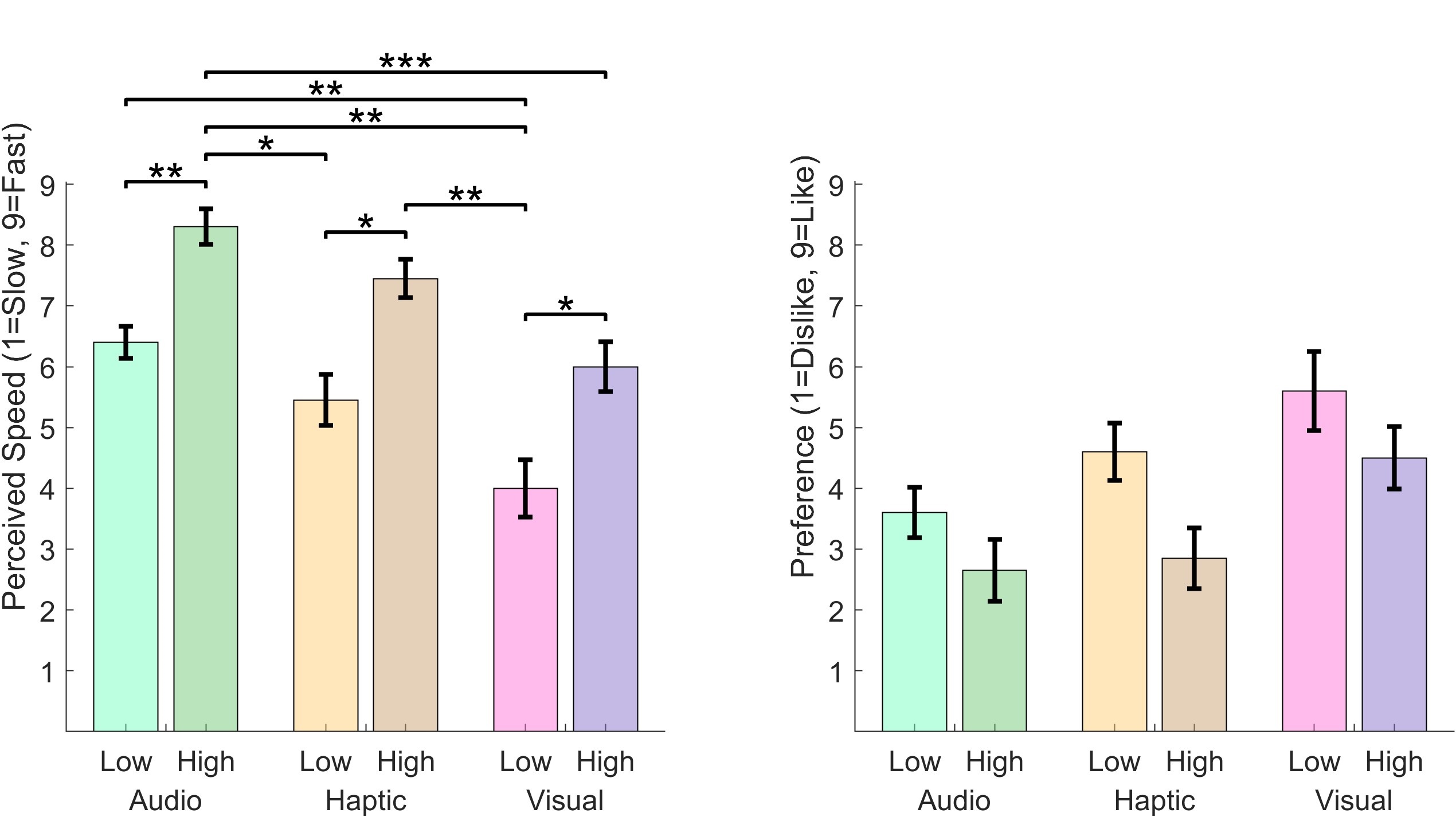}
	\vspace{-0.25in}
	\caption{\revision{Means and standard errors of participant survey responses}. \revision{Statistical significance was determined using Wilcoxon Signed-Rank Tests with p-values indicated with standard significance notation ($*: 0.01 < p < 0.05$, $**: 0.001 < p < 0.01$, $*\!*\!*: p < 0.001$).} Participants were asked how fast they responded to different stimuli (left), and reported that they responded fastest to the high audio stimulus and slowest to the low visual stimulus. Participants were asked to rate each stimulus based on preference (right), and reported that the low visual stimulus was most favored, while the high audio stimulus was least favored. \revision{There were no significant differences between any of the participant survey preferences. } }\label{fig:survey_preference}
     \vspace{-0.15 in}

\end{figure}

\subsection{\revision{Uni-Modal Stimuli
}}

Figure~\ref{fig:2waysingle} shows the response times audio, haptic, and visual stimuli when they are displayed independently at high and low levels. Participants had the shortest response time for the high audio stimulus (405 $\pm$ 50~\revision{ms}) and longest response time for the low visual stimulus (528 $\pm$ 105~\revision{ms}). A two-way repeated measures ANOVA was conducted with type of stimuli (audio, haptic, or visual) and level of stimuli (high/low) as independent factors. The two-way repeated measures ANOVA revealed a main effect of stimulus type ($F(1.33,25.3)=24.1$, $p=\num{1.18E-5}$, $\eta_{p}^{2}=0.559$) and stimulus level ($F(1,19)=59.7$, $p=\num{2.79E-7}$, $\eta_{p}^{2}=0.759$) on response time. These main effects were qualified by an interaction between stimulus type and level ($F(1.65,31.3)=12.5$, $p=\num{2.26E-4}$, $\eta_{p}^{2}=0.398$). 

Because of the significant interaction, the effect of the level of stimulus on response time was analyzed for each type of stimulus using post-hoc pairwise comparisons with Bonferroni correction. These tests found that a higher haptic stimulus resulted in a faster response time than a lower haptic stimulus ($p=\num{1.50E-7}$), a higher audio stimulus resulted in a faster response time than a lower audio stimulus ($p=\num{0.008}$), and a higher visual stimulus resulted in a faster response time than a lower visual stimulus ($p=\num{2.06E-5}$). 

Lastly, the effect of the type of stimulus on response time was analyzed for each level of stimulus. Post-hoc pairwise comparisons with Bonferroni correction showed a significant difference between the low audio and low haptic stimulus ($p=\num{4.62E-7}$) and the low audio and low visual stimulus ($p=\num{1.01E-5}$). There was no significant difference between the low haptic and low visual stimulus. 

Post-hoc testing also showed significant differences between the high audio and high haptic stimulus ($p=\num{2.79E-5}$) and the high audio and high visual stimulus ($p=\num{0.005}$). There was no significant difference between the high haptic and high visual stimulus. 

\subsection{Survey Results}
Participants were asked to fill out a short survey after the experiment to measure their perceived response time, preferred stimulus, and strategies for completing the experiment.

\subsubsection{Perceived Response Time}
Participants were asked to rate their response time on a 9-point Likert scale from 1 (slow) to 9 (fast) for each of the individual stimuli (high audio, low audio, high haptic, low haptic, high visual, low visual). Figure~\ref{fig:survey_preference} shows the reported participant speed for the six individual stimuli. A Friedman's Test was used to identify the effect of stimulus level (low/high) and type (audio/haptic/visual) on perceived response time. There was a significant effect of stimulus type on perceived response time ($\chi^2(2)=36.2, p=\num{1.37E-10}$) and a significant effect of stimulus level on perceived response time ($\chi^2(1)=37.4, p=\num{9.56E-10}$). Wilcoxon Signed-Rank Tests with Bonferroni corrections were used to compare perceived speed, and the adjusted p-values from those tests are shown in Table~\ref{Table: survey_perceived_time}. Participants reported that they believe that they performed fastest for the high audio, and slowest for the low visual stimulus. Participants also reported that they had faster response times for higher stimuli. 

\subsubsection{Preference}

Participants were asked to rate their preference on a 9-point Likert scale from 1 (dislike) to 9 (like) for each of the individual stimuli (high audio, low audio, high haptic, low haptic, high visual, low visual). Figure~\ref{fig:survey_preference} shows the mean preference of each stimulus. A Friedman's Test was used to identify the effect of stimulus level (low/high) and type (audio/haptic/visual) on preference. There was a significant effect of stimulus type on preference ($\chi^2(2)=14.2, p=\num{8.32E-04}$) and a significant effect of stimulus level on perceived response time ($\chi^2(1)=13.2, p=\num{2.76E-04}$). Wilcoxon Signed-Rank Tests with Bonferroni Correction showed no significant differences between any pairings of stimuli. Participants reported that they preferred the low visual stimulus the most and the high audio the least. Across all stimulus types, participants reported higher preference for the lower stimulus level. 



\revision{\subsection{Relationships Between Participant Surveys Response Time}}

\begin{figure}
	\centering
	\includegraphics[width=\columnwidth]{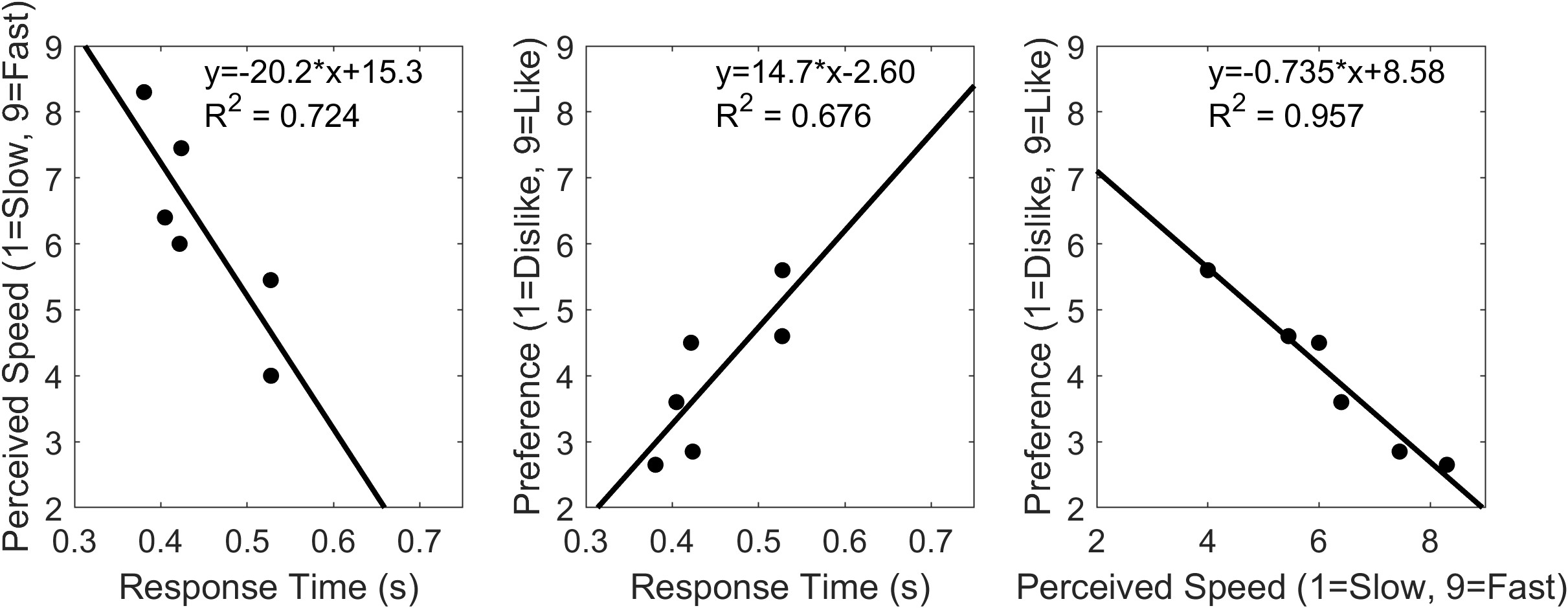}
	\vspace{-0.25in}
	\caption{Relationship between perceived speed and response time (left), preference and response time (middle), and preference and perceived speed (right). The mean response time perceived by participants was related to their actual response time, and participants had a stronger preference towards stimuli with higher response times. Participants had a stronger preference for stimuli to which they felt they responded slower. }\label{fig:rt_vs_survey}
    \vspace{-0.15 in}

\end{figure}

\revision{In Fig.~\ref{fig:rt_vs_survey}, we plotted a linear fit between perceived speed (participant's self-reported response time on a 9-point Likert scale) and measured response time averaged for each of the 6 conditions for which only a single stimulus is played and found that there was a strong relationship ($R^2$ = 0.724). We also found strong relationships between stimuli preference and response time ($R^2$ = 0.676), in addition to stimuli preference and perceived speed ($R^2$ = 0.957). }

\begin{table}[]
\caption{\revision{Adjusted P-values for Uni-modal Response Time Pairwise Comparisons 
}}
\begin{tabular}{ll|rr|rr|rr|}
\cline{3-8}
 &  & \multicolumn{2}{l|}{Audio} & \multicolumn{2}{l|}{Haptic} & \multicolumn{2}{l|}{Visual} \\ \cline{3-8} 
 &  & \multicolumn{1}{l|}{Low} & \multicolumn{1}{l|}{High} & \multicolumn{1}{l|}{Low} & \multicolumn{1}{l|}{High} & \multicolumn{1}{l|}{Low} & \multicolumn{1}{l|}{High} \\ \hline
\multicolumn{1}{|l|}{} & Low & \multicolumn{1}{l|}{\cellcolor[HTML]{9B9B9B}{\color[HTML]{FFFFFF} }} & \textbf{0.009} & \multicolumn{1}{r|}{0.269} & 0.064 & \multicolumn{1}{r|}{\textbf{0.006}} & 1.000 \\ \cline{2-8} 
\multicolumn{1}{|l|}{\multirow{-2}{*}{Audio}} & High & \multicolumn{1}{r|}{\cellcolor[HTML]{9B9B9B}\textbf{}} & \multicolumn{1}{l|}{\cellcolor[HTML]{9B9B9B}{\color[HTML]{FFFFFF} }} & \multicolumn{1}{r|}{\textbf{0.015}} & 0.383 & \multicolumn{1}{r|}{\textbf{0.002}} & \textbf{0.001} \\ \hline
\multicolumn{1}{|l|}{} & Low & \multicolumn{1}{r|}{\cellcolor[HTML]{9B9B9B}} & \cellcolor[HTML]{9B9B9B}\textbf{} & \multicolumn{1}{l|}{\cellcolor[HTML]{9B9B9B}{\color[HTML]{FFFFFF} }} & \textbf{0.011} & \multicolumn{1}{r|}{0.238} & 1.000 \\ \cline{2-8} 
\multicolumn{1}{|l|}{\multirow{-2}{*}{Haptic}} & High & \multicolumn{1}{r|}{\cellcolor[HTML]{9B9B9B}} & \cellcolor[HTML]{9B9B9B} & \multicolumn{1}{r|}{\cellcolor[HTML]{9B9B9B}{\color[HTML]{FFFFFF} \textbf{}}} & \multicolumn{1}{l|}{\cellcolor[HTML]{9B9B9B}{\color[HTML]{FFFFFF} }} & \multicolumn{1}{r|}{\textbf{0.004}} & 0.111 \\ \hline
\multicolumn{1}{|l|}{} & Low & \multicolumn{1}{r|}{\cellcolor[HTML]{9B9B9B}\textbf{}} & \cellcolor[HTML]{9B9B9B}\textbf{} & \multicolumn{1}{r|}{\cellcolor[HTML]{9B9B9B}{\color[HTML]{FFFFFF} }} & \cellcolor[HTML]{9B9B9B}{\color[HTML]{FFFFFF} \textbf{}} & \multicolumn{1}{l|}{\cellcolor[HTML]{9B9B9B}{\color[HTML]{FFFFFF} }} & \textbf{0.033} \\ \cline{2-8} 
\multicolumn{1}{|l|}{\multirow{-2}{*}{Visual}} & High & \multicolumn{1}{r|}{\cellcolor[HTML]{9B9B9B}} & \cellcolor[HTML]{9B9B9B}\textbf{} & \multicolumn{1}{r|}{\cellcolor[HTML]{9B9B9B}{\color[HTML]{FFFFFF} }} & \cellcolor[HTML]{9B9B9B}{\color[HTML]{FFFFFF} } & \multicolumn{1}{r|}{\cellcolor[HTML]{9B9B9B}\textbf{}} & \multicolumn{1}{l|}{\cellcolor[HTML]{9B9B9B}{\color[HTML]{FFFFFF} }} \\ \hline
\end{tabular}
\label{Table: survey_perceived_time}
\end{table}

\vspace{0.2 in}

\section{Discussion}

This study found that a low visual stimulus resulted in the slowest mean response time (mean $\pm$ standard deviation, 528 $\pm$ 105~ms), and a combination of audio, haptic, and visual stimuli at high levels resulted in the fastest mean response time (320 $\pm$ 43~ms). \revision{The response times were slightly faster than other experiments conducted on smartphones by Yoshida and Kiernan et al., which had a mean vibration response time from 662~ms to 825~ms~\cite{yoshida2023cognitive}. Response times in our study may be faster because of a different intensity range and because participants were instructed to respond as fast as possible}. Our results show that when played individually, haptic and visual stimuli had similar response times, with audio being faster. This is different from Chan and Ng~\cite{chan_2012}, who used a different setup and found that a haptic stimulus was the fastest (385~ms), followed by audio (493~ms), then visual (517~ms). Peon and Prattichizzo~\cite{peon_reactiontime} also found that vibrations had a faster response time than audio, which is different compared to our setup. \revision{Deviations from prior literature result from differing stimulus intensities and movement time needed to respond on different platforms~\cite{oldmen_reactiontime}.}

Unlike prior studies that test only a subset of stimuli types or levels~\cite{hecht_reiner_2008, chan_2012, scott2008comparison, belz1999new, jia2019, peon_reactiontime, petermeijer2017, hecht2008multisensory}, we tested 26 combinations of three \revision{modalities }
(audio/haptic/visual) across three intensity levels (high/low/off). \revision{In addition to confirming that a greater number of stimulus types decreases response time~\cite{hecht2008multisensory}, we also found that both the level and the type of stimulus have main effects for modulating response time across most combinations of conditions. This supports the idea that the activation of more sensory modalities with higher intensities may activate a larger neural network for faster processing~\cite{hecht2008multisensory, downar2000nature}.} \revision{People can recognize audio-visual signals} in as little as 45~ms with a processing time of up to 200~ms~\cite{MOLHOLM2002115}. Combined with the time to physically input a response, the full response times in our study match others, with a range of about 200~ms to 1,000~ms~\cite{peon_reactiontime, hecht2006VR, hecht2008multisensory}. 


\revision{Our survey results showed that participants favored visual stimuli over auditory stimuli, but they perceived that they responded fastest to audio stimuli. These relationships may exist, because \revision{certain stimuli} elicit annoyance and a heightened sense of urgency, promoting a faster response~\cite{annoyance}. This also shows that there is a trade-off between response \revision{speed} and stimuli preference that designers of haptic and mobile haptic devices should consider.}

\section{Conclusion}

Our study tested 26 combinations of audio, haptic, and visual stimuli at \revision{three} levels and showed that response time decreases with higher levels of stimuli and more types of stimuli. Our results show that combinations of stimuli affect response times and that people may have a preference for certain stimuli over others. 

In the future, we will broadly distribute our app to measure response time in different populations. In-lab research has shown that factors like education and age impact response time~\cite{chan_2012}, and we would like to further extend this to examine how other factors like physical activity, technology use, socioeconomic status, gender, geography, sleep, and race impacts response time. \revision{Lastly, because response times are unique to specific scenarios and surroundings, we will recruit participants for out-of-lab testing to see how our results change in everyday settings.}

\bibliographystyle{IEEEtran}
\balance

\bibliography{reactionTimes.bib}

\vfill

\end{document}